\documentclass[12pt, letterpaper]{article}

\usepackage{amsmath,amssymb}
\usepackage{cite}
\usepackage{fancyhdr}
\usepackage[top=1in, bottom=1in, left=1in, right=1in]{geometry}
\usepackage[dvipdfm, dvips]{graphicx}
\usepackage{hyperref}

\numberwithin{equation}{section}

\begin{document}


\setcounter{page}{0}
\date{}

\lhead{}\chead{}\rhead{\footnotesize{RUNHETC-2009-30\\SCIPP 09/16\\UCSD-PTH-09-12}}\lfoot{}\cfoot{}\rfoot{}

\title{\textbf{Fixing the Pole in the Pyramid\vspace{0.4cm}}}

\author{Tom Banks$^{1,2}$, Jean-Fran\c{c}ois Fortin$^3$ and Scott Kathrein$^1$\vspace{0.7cm}\\
{\normalsize{$^1$NHETC and Department of Physics and Astronomy, Rutgers University,}}\\
{\normalsize{Piscataway, NJ 08854-8019, USA}}\vspace{0.2cm}\\
{\normalsize{$^2$SCIPP and Department of Physics, University of California,}}\\
{\normalsize{Santa Cruz, CA 95064-1077, USA}}\vspace{0.2cm}\\
{\normalsize{$^3$Department of Physics, University of California,}}\\
{\normalsize{San Diego, CA 92093-0319, USA}}}

\maketitle
\thispagestyle{fancy}

\begin{abstract}
\normalsize
\noindent
We revisit the problem of the hidden sector Landau pole in the Pyramid Scheme.  There is a fixed line in the plane of hidden sector gauge coupling and a Yukawa coupling between the trianon fields.  We postulate that the couplings flow to this line, at a point where the hidden sector gauge coupling is close to the strong coupling edge of its perturbative regime.  Below the masses of the heavier trianons, the model quickly flows to a confining $N_F=N_C=3$ supersymmetric gauge theory, as required by phenomenological considerations.  We study possible discrete R-symmetries, which guarantee, among other things, that the basin of attraction of the fixed line has full co-dimension in the space of R-allowed couplings.  The Yukawa couplings required to get the fixed line violate the pyrma-baryon symmetries we invoked in previous work to find a dark matter candidate.  Omitting one of them, we have a dark matter candidate, and an acceptable RG flow down from the unification scale, if the confinement scale of the hidden sector group is lowered from 5 to 2 TeV.
\end{abstract}


\newpage
\tableofcontents
\vspace{1cm}


\section{Introduction: The Pyramid Scheme and its Pole}\label{sec:intro}

The Pyramid Scheme \cite{Banks:2009rb,Banks:2009nz} is an attractive model of TeV scale physics.  It is a model of direct mediation of supersymmetry (SUSY) breaking, which is consistent with perturbative gauge coupling unification.  Although it contains an interesting scalar pseudo-Nambu Goldstone boson (PNGB), that particle has an MeV scale mass and is consistent with stellar cooling bounds \cite{Banks:2009cx}.  The model explains the absence of all dimension 4 and 5 operators, which could lead to unobserved violations of baryon or lepton number, in terms of a discrete R-symmetry.  It provides a variety of possible dark matter candidates, and might be able to explain the satellite data on cosmic lepton excesses.  It is entirely consistent with the constraints of the theory of Cosmological SUSY Breaking (CSB), with its very low SUSY breaking scale.

In \cite{Banks:2009nz}, two of the authors pointed out a problem with the model.  The hidden sector gauge group must confine at a scale $\Lambda_3$ of order a few TeV.  This is an $SU(3)$ gauge theory with 9 flavors, and is not asymptotically free at high energies.  6 of the flavors have masses above the confinement scale, but these masses are also constrained to be in the multi-TeV range.  Low energy boundary conditions thus appear to lead to a Landau pole in this effective coupling, well below the unification scale.  In \cite{Banks:2009nz} we proposed to solve this problem by assuming that the hidden sector group was in fact $SU(4)$, broken to $SU(3)$ by the Higgs mechanism at a scale of order 50 TeV.  We did not provide a dynamical explanation for the non-zero expectation value, which leads to this Higgs effect.  Furthermore, this assumption forced us to postulate an embarrassingly large discrete R-symmetry.

In this paper we will exorcize the Landau ghost by invoking a famous line of fixed points in the supersymmetric $SU(3)$ gauge theory with 9 flavors, and a certain cubic superpotential \cite{Parkes:1984dh,West:1984dg,Jones:1984cx,Hamidi:1984ft,Hamidi:1984gd,Leigh:1995ep}.  This fixed line is well known to be attractive in the space of gauge/superpotential couplings.  We will show that we can choose our discrete R-symmetry in such a way that it is attractive in all directions in coupling space.  We then postulate initial conditions such that the renormalization group (RG) flow hits the fixed line at the strong coupling end of the perturbative regime.  The gauge coupling remains fixed at this value until the RG scale passes through the masses of the heavier trianons, and six of the flavors decouple.  We then have an $N_F=N_C=3$ gauge theory, with barely perturbative coupling.  This flows rapidly to the confining phase.

This solution to the Landau pole problem has however important consequences for the cosmological implications of the Pyramid Scheme as discussed in \cite{Banks:2009rb}.  Indeed, the extra cubic superpotential breaks explicitly all the accidental symmetries which lead to the light PNGB and stable dark matter candidates.  To keep as many cosmological features as possible we therefore explore the RG flow with only parts of the extra cubic superpotential.  We will show that the tension between Landau poles and cosmological features can be alleviated if the Pyramid strong coupling scale is lowered to a few TeV.

\subsection{The Pyramid Lagrangian}

The Pyramid Scheme is a SUSY gauge theory with gauge group $SU(3)^4\ltimes Z_3$, where the $Z_3$ permutes the last three $SU(3)$ factors (denoted $SU_i(3)$ with $i=1,2,3$) cyclically.  The first $SU_P(3)$ group, the Pyramid group, is the direct mediation sector for this model.  The chiral field content is shown in the following table.
\begin{equation*}
\begin{array}{|c||cccc|}\hline
 & SU_P(3) & SU_1(3) & SU_2(3) & SU_3(3)\\\hline\hline
F^{g}_{12} & 1 & \bar{3} & 3 & 1\\
F^{g}_{23} & 1 & 1 & \bar{3} & 3\\
F^{g}_{31} & 1 & 3 & 1 & \bar{3}\\
{\cal T}_1 & 3 & \bar{3} & 1 & 1\\
\bar{\cal T}_1 & \bar{3} & 3 & 1 & 1\\
{\cal T}_2 & 3 & 1 & \bar{3} & 1\\
\bar{\cal T}_2 & \bar{3} & 1 & 3 & 1\\
{\cal T}_3 & 3 & 1 & 1 & \bar{3}\\
\bar{\cal T}_3 & \bar{3} & 1 & 1 & 3\\
S_{i=1,2,3} & 1 & 1 & 1 & 1\\\hline
\end{array}
\end{equation*}
The index $g$ is three-valued and denotes the standard model generation.  The fields $F^g_{jk}$ contain the standard model generations, plus 36 extra fields, which are vector-like under the standard model.  This includes 3 candidate pairs of Higgs fields.  We will continue to follow the procedure in \cite{Banks:2009rb,Banks:2009nz} and work solely at scales below the unification scale.  In the effective Lagrangian at those scales, we do not have to make a commitment to whether the Higgs fields originate from these multiplets or from another source.

A few remarks about GUT scale physics are however in order.  We note for example that the standard model field content of this model arises naturally from 3 D3 branes at the $Z_3$ orbifold, and that the fields $F^g_{12}$ can have VEVs which break the base of the Pyramid gauge group down to the standard model.  One major advantage of trinification over conventional grand unification, is that group theory does not allow couplings of the $F^g_{ij}$ fields to the gauge field strengths, which could lead to order one modifications of gauge coupling relations if the scale of these terms were the same as the unification scale.  All such couplings necessarily involve all 3 of the $F^g_{ij}$ fields and their contribution to tree level coupling renormalization vanishes, because $F^{g}_{23,31}$ have vanishing VEVs.

A plausible scenario is that the D3 branes are embedded in a compact 6-fold and that the large volume of this 6-fold in 10 dimensional Planck units is responsible for the ratio between the four dimensional Planck and unification scales.  Powers of the 6-fold K\"{a}hler moduli are the explanation of the texture of the quark and lepton mass matrices and the ratio between the neutrino see-saw scale and the unification scale.  We will now write the superpotential for the low energy theory below the unification scale.  It consists of two parts.  The first of these survives in the limit that the cosmological constant (c.c.) is taken to zero.  It preserves a discrete R-symmetry and an ordinary discrete symmetry, and is the most general function of the fields compatible with these symmetries,
\begin{equation}
W_{\Lambda=0}=\sum_{i,j}y_{ij}S_i{\rm tr}({\cal T}_j\bar{\cal T}_j)+\sum_i[u_i{\rm det}({\cal T}_i)+\bar{u}_i{\rm det}(\bar{\cal T}_i)+\beta_iS_iH_uH_d]+W_{\rm std}.
\end{equation}
$W_{\rm std}$ is the conventional superpotential of the MSSM, which we assume is determined by unification scale physics.  We will explore the possible choices of discrete groups in section \ref{sec:sym}.  There will also be higher dimension operators, suppressed by powers of the ten dimensional Planck mass.  These will lift all the flat directions in the renormalizable superpotential we have written, and freeze the system at an R-symmetric and supersymmetric point.  Note that, as we will discuss in the next section, we need to invoke an ordinary discrete symmetry as well as the fundamental R-symmetry in order to forbid all trilinear terms in the $S$ fields.

According to the hypothesis of Cosmological SUSY Breaking (CSB) the theory of a future asymptotically de Sitter (dS) universe is a finite quantum system \cite{Banks:2000fe}.  The logarithm of the total number of states in the system is $\pi(RM_P)^2$.  Of these, of order $(RM_P)^{3/2} $ can be viewed as states of particles in a single horizon volume, while the others are, from the point of view of a local observer, confined to the horizon.  The horizon states form a thermal bath, with temperature $T_{\rm dS}=\frac{1}{2\pi R}$ for the particle states.  In the limit of vanishing c.c., the horizon states decouple and the particles become exactly super-Poincar\'e invariant.  When the c.c. is very small, we can describe the system, approximately, by $\mathcal{N}=1$ SUGRA with spontaneously broken SUSY.  In the low energy effective field theory, the terms which induce spontaneous SUSY breaking are put in by hand.  They break the R-symmetry, which guarantees Poincar\'e invariance in the zero c.c. limit, explicitly.  Heuristically \cite{Banks:2002wj} one can view them as coming from diagrams in which a single gravitino propagates out to the horizon, interacts with the degrees of freedom there and propagates back.  There is no reason to expect these terms to satisfy the usual genericity rules of effective field theory.  They break R-symmetry because the gravitino R-charge is lost to the horizon.

At the present time, the only rules that we know that constrain these terms are
\begin{itemize}
\item They must enforce the relation
\begin{equation}
m_{3/2}=K\Lambda^{1/4}.
\end{equation}
Recent estimates \cite{Banks:2009rb} indicate that $K\sim10$.
\item The Coleman de Lucia tunneling rate computed in the effective field theory must be of order $e^{-\pi(RM_P)^2}$, so that the decay of the dS vacuum state can be interpreted as a Heisenberg/Poincar\'e recurrence rather than an instability.  As argued in \cite{Banks:2009nz}, this means that the renormalizable effective field theory cannot have a supersymmetric ground state.  Recalling that R-symmetry is explicitly broken, and invoking an old result of Nelson and Seiberg \cite{Nelson:1993nf}, we conclude that the R-breaking terms {\it must} be non-generic.
\end{itemize}

We postulate that the requisite R-breaking terms are
\begin{equation}
\delta W_{\Lambda\neq0}=W_0+\sum_i[m_i{\rm tr}({\cal T}_i\bar{\cal T}_i)+\mu_i^2S_i]+\mu H_uH_d.
\end{equation}
The parameters $W_0$ and $\mu_i^2$ are all necessary to enforce the above rules, while $m_i$ and $\mu$ are inserted for frankly phenomenological reasons.  All of these parameters should vanish with the c.c..  For example $\mu_i^2\sim\Lambda^{1/4}m_P$, while $W_0$ is of order $\Lambda^{1/4}m_P^2$ and is fine tuned to give an effective field theory value for the c.c. consistent with its value in the fundamental theory.  Roughly speaking we expect $\mu\sim m_i\sim\mu_i$, but we have no clean estimate of the prefactors in these relations.  We will be guided by phenomenology in our choices of the actual numbers.

We note two features of this formula which are new compared to previous treatments of this subject.  First of all, we have realized that there is no good reason to omit the $\mu$ term from the R-breaking part of the effective Lagrangian.  In previous work we have tried to realize the phenomenologically necessary $\mu$ term by the typical dynamical mechanism invoked in the NMSSM.  However, we have realized that there is no reason that it could not arise from interactions with the horizon, since its absence was due to the R-symmetry.  This hypothesis ameliorates the $\mu$ problem of many gauge mediated models, as well as the little hierarchy problem.  As we will review in a moment, our model has $\tan\beta\sim1$ and several Yukawa couplings of singlets to the Higgs.  There is no problem pushing the Higgs mass above current experimental bounds.  When this mechanism is exploited in the NMSSM, one has a problem generating a $\mu$ term that is large enough, because the singlet VEVs vanish at the minimum of the potential.  Here $\mu$ is a free parameter coming from mechanisms that cannot be understood in effective field theory.  Field theorists will scoff that this is no solution at all, but we have begun to realize that, within the framework of CSB, it is not only the cosmological constant problem whose solution will look strange to a doctrinaire field theorist.

Our second remark is of a similar nature, but somewhat more complete.  We claim that there is an argument, within the framework of CSB, that the phases of the R-breaking parameters $\mu$, $m_i$ and $\mu_i$  are all very small.  Using the prescription of \cite{Banks:2002wj} these parameters are computed in terms of Feynman diagrams in low energy effective field theory for vanishing c.c. except that the gravitino mass in the lines that go out to the horizon is kept non-zero and computed self consistently.  Furthermore the system on the horizon is thermal, with a very large near horizon temperature.  Thus, if there is any sense in which the CP violation in the low energy field theory is spontaneous, with any scale of spontaneous breaking, the system on the horizon is CP invariant.

It is easy to see that the R-preserving part of the theory is CP invariant, except for the usual CKM phase.  We can rotate away all Yukawa phases and gauge theory $\theta$ parameters, with Peccei-Quinn like transformations.  Normally this leads to either massless fermions or axion fields, but the symmetries are broken by $\delta W_{\Lambda\neq0}$.  In standard field theoretic calculations one would take these symmetry breaking terms to be generic because there is no reason for the new physics responsible for them to be CP invariant.  However, in the Feynman diagrams of CSB, the only source of CP violation comes from standard model loop corrections to the basic gravitino diagram.  The external lines of the diagram are at low momentum, so the CP violating corrections are small.  This is a novel solution of the strong CP problem \cite{Banks:2009hx}.


\section{Symmetries}\label{sec:sym}

We find in this section the allowed discrete symmetry groups that obey all anomaly constraints, rule out all dangerous baryon or lepton number violating operators, allow the neutrino seesaw term, rule out the $\mu$ term, and rule out trilinear terms in the $S$ fields while allowing the trilinear couplings of the trianon fields.

The appropriate discrete R-symmetry must be exact in the zero c.c. limit where SUSY is restored, and only approximate when the c.c. is non-zero and SUSY is broken.  Moreover, in order for the discrete R-symmetry to be anomaly free, the 't Hooft operators coming from standard model and $SU_P(3)$ instantons must vanish.  The discrete R-symmetry therefore imposes tight constraints on the R-charges of the different fields since only specific terms are allowed in the superpotential and the K\"{a}hler potential.  To simplify the investigation, in this section the R-charges of the fields are denoted by the fields themselves and all equations have to be satisfied only modulo $N$ due to the discrete $Z_N$ R-symmetry.  Apart from standard model fields, the extra matter fields in GUT notation are shown in the following table.
\begin{equation*}
\begin{array}{|c||cccc|}\hline
  & SU_P(3) & SU_1(3) & SU_2(3) & SU_3(3)\\\hline\hline
{\cal T}_1 & 3 & \bar{3} & 1 & 1\\
\bar{\cal T}_1 & \bar{3} & 3 & 1 & 1\\
{\cal T}_2 & 3 & 1 & \bar{3} & 1\\
\bar{\cal T}_2 & \bar{3} & 1 & 3 & 1\\
{\cal T}_3 & 3 & 1 & 1 & \bar{3}\\
\bar{\cal T}_3 & \bar{3} & 1 & 1 & 3\\
S_{i=1,2,3} & 1 & 1 & 1 & 1\\\hline
\end{array}
\end{equation*}
The only superpotential terms which are required at the renormalizable level in the zero c.c. limit are
\begin{equation*}
W_{\Lambda=0}\supset
S_i{\rm tr}({\cal T}_j\bar{\cal T}_j),\;S_iH_uH_d,\;H_uQ\bar{U},\;H_dQ\bar{D},\;H_dL\bar{E},\;\{{\rm det}({\cal T}_i),\;{\rm det}(\bar{\cal T}_i),\;(LH_u)^2\}
\end{equation*}
which implies that the R-charges satisfy
\begin{eqnarray*}
{\cal T}_i+\bar{\cal T}_i &=& 2-S\\
H_u &=& 2-H_d-S\\
\bar{U} &=& H_d+S-Q\\
\bar{D} &=& 2-H_d-Q\\
\bar{E} &=& 2-H_d-L.
\end{eqnarray*}
where the gaugino R-charges are assumed canonical.  Here the extra relations coming from the terms in brackets are not taken into account yet.  For example the baryon operators force $S_i={\cal T}_i=\bar{\cal T}_i$.  Notice moreover that the seesaw operator must be allowed in order to give neutrino masses.  By naturalness of the low energy effective field theory in the zero c.c. limit, all remaining renormalizable superpotential terms must be forbidden by the discrete R-symmetry, or by a discrete ordinary symmetry.  Moreover, dangerous dimension four and five $B$ and $L$ violating terms must be forbidden as well by the discrete symmetries to insure proton stability on appropriate timescales.

The vanishing of the 't Hooft operators implies the vanishing of the following equations
\begin{eqnarray*}
SU_P(3)^2U_R(1) &\Rightarrow& 2\cdot3+3({\cal T}_1+\bar{\cal T}_1+{\cal T}_2+\bar{\cal T}_2+{\cal T}_3+\bar{\cal T}_3-6)=6-9S\\
SU_C(3)^2U_R(1) &\Rightarrow& 2\cdot3+6(Q-1)+3(\bar{U}+\bar{D}-2)+3({\cal T}_3+\bar{\cal T}_3-2)=0\\
SU_L(2)^2U_R(1) &\Rightarrow& 2\cdot2+(H_u+H_d-2)+9(Q-1)+3(L-1)\\
 && \hspace{1cm}+3({\cal T}_2+\bar{\cal T}_2-2)=3(3Q+L)-4(2-S).
\end{eqnarray*}
The vanishing of the 't Hooft operator coming from $SU_C(3)$ standard model instantons does not constraint the R-charges.  The remaining equations lead to $S=22-6(3Q+L)$ with the 't Hooft constraints $27(3Q+L)-96=0$.

The forbidden (non-$B$ and non-$L$ violating) renormalizable superpotential terms can be combined into $4$ groups,
\begin{eqnarray*}
G_1^{\rm (ren)}=\{{\rm tr}({\cal T}_i\bar{\cal T}_i),\;H_uH_d\} &\Rightarrow& S\\
G_2^{\rm (ren)}=\{S\} &\Rightarrow& S-2\\
G_3^{\rm (ren)}=\{S^2\} &\Rightarrow& 2S-2\\
G_4^{\rm (ren)}=\{S^3\} &\Rightarrow& 3S-2.
\end{eqnarray*}
Moreover, the dangerous renormalizable and higher-dimensional $B$ and $L$ violating superpotential and K\"{a}hler potential terms can be combined into $9$ groups,
\begin{eqnarray*}
G_1^{\rm (\not{B}\,or\,\not{L})}=\{LL\bar{E},\;LQ\bar{D},\;SLH_u\} &\Rightarrow& L-H_d\\
G_2^{\rm (\not{B}\,or\,\not{L})}=\{\bar{U}\bar{D}\bar{D}\} &\Rightarrow& 3Q+H_d-S-2\\
G_3^{\rm (\not{B}\,or\,\not{L})}=\{LH_u,\;Q\bar{U}\bar{E}H_d,\;\bar{U}\bar{D}^*\bar{E},\;H_u^*H_d\bar{E},\;Q\bar{U}L^*\} &\Rightarrow& L-H_d-S\\
G_4^{\rm (\not{B}\,or\,\not{L})}=\{QQQL\} &\Rightarrow& 3Q+L-2\\
G_5^{\rm (\not{B}\,or\,\not{L})}=\{QQQH_d,\;QQ\bar{D}^*\} &\Rightarrow& 3Q+H_d-2\\
G_6^{\rm (\not{B}\,or\,\not{L})}=\{\bar{U}\bar{U}\bar{D}\bar{E}\} &\Rightarrow& 3Q+L-2S-2\\
G_7^{\rm (\not{B}\,or\,\not{L})}=\{LH_uH_dH_u\} &\Rightarrow& L-H_d-2S+2\\
G_8^{\rm (\not{B}\,or\,\not{L})}=\{SLL\bar{E},\;SLQ\bar{D},\;S^2LH_u\} &\Rightarrow& L-H_d+S\\
G_9^{\rm (\not{B}\,or\,\not{L})}=\{S\bar{U}\bar{D}\bar{D}\} &\Rightarrow& 3Q+H_d-2S-2.
\end{eqnarray*}
All operators belonging to the same group share the same R-charge.

Taking into account all the relations and constraints, it is possible to engineer the following superpotential of the low energy effective theory in the zero c.c. limit,
\begin{multline}
W_{\Lambda=0}=\sum_{i,j}y_{ij}S_i{\rm tr}({\cal T}_j\bar{\cal T}_j)+\sum_i\left[u_i{\rm det}({\cal T}_i)+\bar{u}_i{\rm det}(\bar{\cal T}_i)+\beta_iS_iH_uH_d\right]\\
+\lambda_uH_uQ\bar{U}+\lambda_d H_dQ\bar{D}+\lambda_LH_dL\bar{E}+\frac{\lambda_{\nu}}{m_P}(L H_u)^2
\end{multline}
where all allowed renormalizable terms are present and all dangerous terms are forbidden by a discrete $Z_8$ R-symmetry with $S_i={\cal T}_i=\bar{\cal T}_i=6$, $Q=5$, $L=1$ and $H_d=0$.  Notice that, in order for the theory to exhibit SUSY breaking in stable states, the trilinear $S^3$ terms, which would be allowed by the discrete R-symmetry, must be forbidden by an extra ordinary discrete symmetry.  This extra ordinary discrete symmetry is also broken explicitly by the R-breaking operators.

This generic superpotential is supplemented by the non-generic R-symmetry breaking superpotential induced by interactions with the horizon
\begin{equation}
\delta W_{\Lambda\neq0}=W_0+\sum_i\left[m_i{\rm tr}({\cal T}_i\bar{\cal T}_i)+\mu_i^2S_i\right]+\mu H_uH_d
\end{equation}
when the c.c. is turned on.  The latter does not follow the rules of effective field theory since it comes from interactions with the cosmological horizon.


\section{Renormalization Group Flow}\label{sec:RG}

In  this section, the renormalization group behavior is analyzed to determine if an appropriate value of the $SU_P(3)$ confinement scale $\Lambda_3$ can be obtained while also keeping all parameters perturbative up to the GUT scale.  We remember that $\Lambda_3\sim5$ TeV in order to have a chargino mass consistent with experiment and a gravitino mass consistent with CSB.  The full two-loop beta functions are used (see Appendix \ref{app:RG} for a list of beta functions to lowest non-trivial order), along with zeroth order threshold effects for each of the trianons (${\cal T}_1$ is integrated out at $m_1=15$ TeV, ${\cal T}_3$ at $m_3=12$ TeV, and ${\cal T}_2$ at $m_2=9$ TeV).  The location of $\Lambda_3$ is determined by evolving the RG equations with initial conditions specified at the unification scale, and the location of the Landau pole determined by specifying initial conditions at the weak scale.  Due to the IR attractive fixed line behavior of the theory with pyrma-baryon operators ${\cal T}_i{\cal T}_i{\cal T}_i$, the latter turn out to have a large effect, and the results will be divided here based on how many of these operators are allowed in the Lagrangian.  Moreover, since the Pyramid group gauge coupling at $m_1$ is preferred close to the strong coupling end of the perturbative regime, Landau poles are present at very low scale if there is no fixed line.

It is important to mention the following caveat: when the coupling constant for the third pyrma-baryon operator is allowed to be non-zero, the color gauge coupling running is affected at two loops.  This destroys standard model gauge coupling unification by as much as 15\%.  Note however, that one-loop gauge coupling running is always unaffected here, and that some threshold effects are being ignored at both the masses of the trianons and at the unification scale.  Because of their smaller values, the other two standard model gauge couplings are not significantly affected by the non-zero values of the pyrma-baryon coupling constants.  In the cases where one or more of the pyrma-baryon Yukawa couplings is turned off, we can choose one of the vanishing couplings to be the third one, eliminating this potential problem.  However, it is hard to judge what the preferred scenario is without a full calculation of threshold effects, including those at the unification scale.

\subsection{Zero or One Allowed Pyrma-baryon Couplings}

In these cases, if we require that the system stay perturbative up to the GUT scale, we find that $\Lambda_3$ is at most of order 300 GeV or 900 GeV, with zero or one baryon couplings respectively.  Conversely, when $\Lambda_3\sim5$ TeV, the $SU_P(3)$ gauge coupling has a Landau pole near 300 -- 400 TeV.  Hence, dynamics at this intermediate scale would have to be introduced in order to obtain a phenomenologically viable value of $\Lambda_3$.  However, in this case two or three of the accidental pyrma-baryon number symmetries are still present, which leads to an attractive dark matter candidate scenario which may explain data from the ATIC, FERMI and PAMELA experiments, as well as a potential explanation of baryogenesis, as discussed in \cite{Banks:2009rb}.

\subsection{Two Allowed Pyrma-baryon Couplings}

In this particular case, if we require that the system stay perturbative up to the GUT scale, we find that $\Lambda_3$ is at most of order 2 TeV.  When $\Lambda_3\sim 5$ TeV, the $SU_P(3)$ gauge coupling has a Landau pole near 1500 TeV.  Hence, we again find that dynamics at an intermediate scale are required to satisfy our initial estimates on the location of the $SU_P(3)$ confinement scale.  With only one of the pyrma-baryon number symmetries still present, the attractive dark matter scenario discussed in \cite{Banks:2009rb} is lost.  However, the theory still has a dark matter candidate, the lightest pyrma-baryon carrying the conserved quantum number.  This must be either ${\cal T}_{1,3}$, since the low energy dynamics spontaneously breaks the second pyrma-baryon number.  Note that if we insist that the third pyrma-baryon number is conserved we also remove the two loop correction to standard model coupling unification.  This scenario also requires a primordial asymmetry in the conserved baryon number, which is tuned to obtain the right dark matter density.  The MeV scale PNGB of the original Pyramid scheme is lifted, since there is no longer an approximate symmetry.

\subsection{All Pyrma-baryon Couplings}

With all pyrma-baryon operators present, we may exploit the IR attractive fixed line known to exist in $SU(3)$ SUSY gauge theory with 9 flavors \cite{Parkes:1984dh,West:1984dg,Jones:1984cx,Hamidi:1984ft,Hamidi:1984gd,Leigh:1995ep}.  In the absence of other couplings, the fixed line is along $g_P^2=\frac{3}{4}u^2$, where $u$ is the coefficient of each of the pyrma-baryon operators.  While its exact location is disturbed slightly by the rest of the theory, it is still attractive.  By specifying barely perturbative initial conditions at the GUT scale for $u_i$ as well as the $SU_P(3)$ gauge coupling, we have $\Lambda_3\sim5$ TeV without any dynamics at an additional scale.  However, with all the pyrma-baryon operators present, none of the pyrma-baryon number preserving symmetries exist, and the theory has no dark matter candidate.


\section{Conclusion}\label{sec:conc}

While the fixed line is the most elegant solution of the Landau pole problem in the hidden sector of the Pyramid Scheme, the model in which the ${\cal T}_3^3$ couplings are absent seems to hold the most phenomenological promise.  This model has a dark matter candidate, though it no longer provides a possible explanation for the various lepton excesses and the WMAP haze.  The MeV scale PNGB whose decay leads to those lepton excesses is lifted by the terms we introduce to obtain an acceptable RG flow.

By lowering the confinement scale $\Lambda_3$, this version of the Pyramid Scheme also increases the tension between the model and the lower bounds on chargino masses.  The calculation of chargino masses involves unknown coefficients coming from the strong dynamics of $SU_P (3)$ below $\Lambda_3$ and those coefficients must now be assumed larger by a factor of $2$ than in previous work.  Detailed verification of the Pyramid Scheme will require a much more quantitative solution of strongly coupled SUSY gauge theories than we have at present.

Although it is not germane to the main thrust of this article we also want to emphasize two new changes in point of view that we noted in the introduction.  We no longer seek a conventional dynamical explanation for the $\mu$ term of the MSSM, even though our model contains fields like the singlet of the NMSSM, whose VEV could explain the $\mu$ term.  In fact, this term violates the fundamental R-symmetry of the limiting version of the theory with vanishing c.c..  There is every reason to assume that it arises through interactions with the horizon.  While this explanation will be anathema to the dogmatic effective field theorist, it completely resolves the tensions known in the literature as {\it the little hierarchy problem} and the {\it $\mu-B_\mu$ problem}\footnote{To be fair, it merely removes them from the domain of effective field theory, and requires they be understood in terms of a detailed theory of interactions of the standard model with degrees of freedom on the horizon.}.

We also provided an argument that the phases of the R-violating terms are very small.  This provides a completely new explanation of the strong CP problem, with neither an axion nor a massless quark.  It requires only that CP violation be spontaneous in the very general sense that it goes away in the very high temperature environment near the cosmological horizon.

In previous work on TeV scale models compatible with the ideas of CSB, TB has tried to hew as closely as possible to the tenets of effective field theory.  Interactions with the horizon were invoked only to explain the relation between the value of the c.c. and the scale of SUSY breaking.  In point of fact, this was a bit disingenuous, because one also had to assume that the R-violating terms coming from the horizon did not include terms incompatible with proton stability.  In our current view, the nature of interactions with the horizon will eventually be seen to be the explanation of the relation between SUSY breaking and the c.c., approximate $B$ and $L$ conservation, CP invariance of the strong interactions, the little hierarchy problem, the origin of the $\mu$ term, and the $\mu-B_\mu$ problem.

We are beginning to reach the limits of what can be done in this field using only low energy field theory techniques.  On that side of the problem, the most pressing issue would seem to be to develop techniques for quantitative solution of strongly coupled SUSY QCD.  The two other avenues for progress are an attempt to relate the Pyramid Scheme to conventional string theoretic approaches to unification, and a return to the fundamental description of CSB in terms of holographic space-time.  The general outline of the Pyramid Scheme fits nicely into a local model including D3 branes at a $Z_3$ orbifold point, and nearby D7 branes.  A compact model would be a form of F-theory, with a singular 3-fold base.  Preliminary analysis \cite{F} suggests that the crucial questions involve possible terms in the superpotential, which are power law in the K\"{a}hler moduli of the base.  It is not clear how to determine whether such terms exist.  Further work on all these issues is in progress.

\subsection*{Acknowledgments}

We would like to thank M. Cvetic, G. Moore, and especially D.E. Diaconescu for valuable discussions.  The research of TB and SK was supported in part by DOE grant DE-FG03-92ER40689.  The research of JFF was supported in part by DOE grant DOE-FG03-97ER40546.


\appendix

\section{IR Attractive Fixed Line Behavior}\label{app:RG}

In order for perturbative gauge coupling unification to occur, the $SU_P(3)$ theory with 9 flavors, i.e. the theory at a scale $\mu$ where $\max\{m_1,m_2,m_3\}<\mu<M_{\rm GUT}$, must exhibit the IR attractive fixed line behavior of $\mathcal{N}=1$ $SU(N)$ SQCD with $3N$ flavors and baryon operators \cite{Parkes:1984dh,West:1984dg,Jones:1984cx,Hamidi:1984ft,Hamidi:1984gd,Leigh:1995ep} in some phenomenologically-viable part of parameter space.  Under the relevant $SU_P(3)\times SU_C(3)\times SU_L(2)\times U_Y(1)$ gauge group, the extra matter fields decompose as shown in the following table.
\begin{equation*}
\begin{array}{|c||c|}\hline
{\rm Fields} & SU_P(3)\times SU_C(3)\times SU_L(2)\times U_Y(1)\\\hline\hline
{\cal T}_1+\bar{\cal T}_1 & (\bar{3},1,1)_{-2/3}\oplus2\cdot(\bar{3},1,1)_{1/3}+{\rm h.c}\\
{\cal T}_2+\bar{\cal T}_2 & (\bar{3},1,2)_{1/6}\oplus(\bar{3},1,1)_{-1/3}+{\rm h.c}\\
{\cal T}_3+\bar{\cal T}_3 & (\bar{3},3,1)_0\oplus(3,\bar{3},1)_0\\
S_{i=1,2,3} & (1,1,1)_0\\\hline
\end{array}
\end{equation*}
The full superpotential also decomposes and becomes
\begin{eqnarray*}
W &=& \sum_i\left[y_{i3}S_i{\rm tr}({\cal T}_3\bar{\cal T}_3)+y_{i2}^{\rm d}S_i{\rm tr}({\cal T}_2^{\rm d}\bar{\cal T}_2^{\rm d})+y_{i2}^{\rm s}S_i{\rm tr}({\cal T}_2^{\rm s}\bar{\cal T}_2^{\rm s})+\sum_ky_{i1}^kS_i{\rm tr}({\cal T}_1^k\bar{\cal T}_1^k)\right.\\
 && \left.+u_i{\rm det}({\cal T}_i)+\bar{u}_i{\rm det}(\bar{\cal T}_i)+\beta_iS_iH_uH_d+\mu_i^2S_i\right]\\
 && +m_3{\rm tr}({\cal T}_3\bar{\cal T}_3)+m_2^{\rm d}{\rm tr}({\cal T}_2^{\rm d}\bar{\cal T}_2^{\rm d})+m_2^{\rm s}{\rm tr}({\cal T}_2^{\rm s}\bar{\cal T}_2^{\rm s})+\sum_km_1^k{\rm tr}({\cal T}_1^k\bar{\cal T}_1^k)+W_0\\
 && +\mu H_uH_d+\sum_{i,j}\left[(\lambda_u)_{ij}H_uQ_j\bar{U}_i+(\lambda_d)_{ij}H_dQ_j\bar{D}_i+(\lambda_L)_{ij}H_dL_j\bar{E}_i\right]+\frac{\lambda_{\nu}}{m_P}(L H_u)^2
\end{eqnarray*}
where ${\cal T}_2^{\rm d}$ and ${\cal T}_2^{\rm s}$ are the $SU_L(2)$ doublet and singlet respectively.

To lowest non-trivial order, the $\beta$-functions for the gauge couplings are \cite{Martin:1993zk}
\begin{equation}
\frac{dg_a}{dt}=\frac{1}{16\pi^2}\beta_{g_a}^{(1)}+\frac{1}{(16\pi^2)^2}\beta_{g_a}^{(2)}
\end{equation}
where $t=\ln(\mu/\mu_0)$ and
\begin{eqnarray*}
\beta_{g_1}^{(1)} &=& \frac{3}{5}g_1^3(11+4+1+0)\\
\beta_{g_2}^{(1)} &=& g_2^3(1+0+3+0)\\
\beta_{g_3}^{(1)} &=& g_3^3(-3+0+0+3)\\
\beta_{g_P}^{(1)} &=& g_P^3(-9+3+3+3).
\end{eqnarray*}
The contributions are ordered as follows: $({\rm MSSM}+{\cal T}_1+{\cal T}_2+{\cal T}_3)$.  The two-loop $\beta$-functions are needed for both $SU(3)$ gauge groups since their one-loop $\beta$-functions vanish in a certain regime,
\begin{eqnarray*}
\beta_{g_3}^{(2)} &=& g_3^3[g_1^2(11/5+0+0+0)+g_2^2(9+0+0+0)+g_3^2(14+0+0+34)\\
 && +g_P^2(0+0+0+16)]-g_3^3\left[4{\rm tr}(\lambda_u^\dagger\lambda_u)+4{\rm tr}(\lambda_d^\dagger\lambda_d)+6\sum_i|y_{i3}|^2+6|u_3|^2+6|\bar{u}_3|^2\right]\\
\beta_{g_P}^{(2)} &=& g_P^3[2g_1^2+6g_2^2+16g_3^2+48g_P^2]\\
 && -g_P^3\sum_i\left[6|y_{i3}|^2+4|y_{i2}^{\rm d}|^2+2|y_{i2}^{\rm s}|^2+2\sum_k|y_{i1}^k|^2+6|u_i|^2+6|\bar{u}_i|^2\right].
\end{eqnarray*}
The anomalous dimensions of the extra matter fields are
\begin{eqnarray}
\gamma_{{\cal T}_3}^{\,\,\,{\cal T}_3} &=& \sum_i|y_{i3}|^2+2|u_3|^2-\left[\frac{8}{3}g_3^2+\frac{8}{3}g_P^2\right]\\
\gamma_{{\cal T}_2^{\rm d}}^{\,\,\,{\cal T}_2^{\rm d}} &=& \sum_i|y_{i2}^{\rm d}|^2+2|u_2|^2-\left[\frac{1}{30}g_1^2+\frac{3}{2}g_2^2+\frac{8}{3}g_P^2\right]\\
\gamma_{{\cal T}_2^{\rm s}}^{\,\,\,{\cal T}_2^{\rm s}} &=& \sum_i|y_{i2}^{\rm s}|^2+2|u_2|^2-\left[\frac{2}{15}g_1^2+\frac{8}{3}g_P^2\right]\\
\gamma_{{\cal T}_1^1}^{\,\,\,{\cal T}_1^1} &=& \sum_i|y_{i1}^1|^2+2|u_1|^2-\left[\frac{8}{15}g_1^2+\frac{8}{3}g_P^2\right]\\
\gamma_{{\cal T}_1^2}^{\,\,\,{\cal T}_1^2} &=& \sum_i|y_{i1}^2|^2+2|u_1|^2-\left[\frac{2}{15}g_1^2+\frac{8}{3}g_P^2\right]\\
\gamma_{{\cal T}_1^3}^{\,\,\,{\cal T}_1^3} &=& \sum_i|y_{i1}^3|^2+2|u_1|^2-\left[\frac{2}{15}g_1^2+\frac{8}{3}g_P^2\right]\\
\gamma_{S_i}^{\,\,\,S_j} &=& 9y_{i3}^*y_{j3}+6y_{i2}^{{\rm d}*}y_{j2}^{\rm d}+3y_{i2}^{{\rm s}*}y_{j2}^{\rm s}+3\sum_ky_{i1}^{k*}y_{j1}^k+2\beta_i^*\beta_j
\end{eqnarray}
while the anomalous dimensions of the standard model fields are
\begin{eqnarray}
\gamma_{H_u}^{\,\,\,H_u} &=& \sum_i|\beta_i|^2+3{\rm tr}(\lambda_u^\dagger\lambda_u)-\left[\frac{3}{10}g_1^2+\frac{3}{2}g_2^2\right]\\
\gamma_{H_d}^{\,\,\,H_d} &=& \sum_i|\beta_i|^2+3{\rm tr}(\lambda_d^\dagger\lambda_d)+{\rm tr}(\lambda_L^\dagger\lambda_L)-\left[\frac{3}{10}g_1^2+\frac{3}{2}g_2^2\right]\\
\gamma_{Q_i}^{\,\,\,Q_j} &=& (\lambda_u^\dagger\lambda_u)_{ij}+(\lambda_d^\dagger\lambda_d)_{ij}-\left[\frac{1}{30}g_1^2+\frac{3}{2}g_2^2+\frac{8}{3}g_3^2\right]\delta_{ij}\\
\gamma_{\bar{U}_i}^{\,\,\,\bar{U}_j} &=& 2(\lambda_u\lambda_u^\dagger)_{ji}-\left[\frac{8}{15}g_1^2+\frac{8}{3}g_3^2\right]\delta_{ij}\\
\gamma_{\bar{D}_i}^{\,\,\,\bar{D}_j} &=& 2(\lambda_d\lambda_d^\dagger)_{ji}-\left[\frac{2}{15}g_1^2+\frac{8}{3}g_3^2\right]\delta_{ij}\\
\gamma_{L_i}^{\,\,\,L_j} &=& (\lambda_L^\dagger\lambda_L)_{ij}-\left[\frac{3}{10}g_1^2+\frac{3}{2}g_2^2\right]\delta_{ij}\\
\gamma_{\bar{E}_i}^{\,\,\,\bar{E}_j} &=& 2(\lambda_L\lambda_L^\dagger)_{ji}-\frac{6}{5}g_1^2\delta_{ij}.
\end{eqnarray}
Therefore, to lowest non-trivial order, the $\beta$-functions for the superpotential couplings are given by
\begin{eqnarray}
16\pi^2\frac{dy_{ij}^k}{dt} &=& y_{ij}^k\gamma_{{\cal T}_j^k}^{\,\,\,{\cal T}_j^k}+y_{ij}^k\gamma_{\bar{\cal T}_j^k}^{\,\,\,\bar{\cal T}_j^k}+\sum_my_{mj}^k\gamma_{S_m}^{\,\,\,S_i}\\
16\pi^2\frac{du_3}{dt} &=& 3u_3\gamma_{{\cal T}_3}^{\,\,\,{\cal T}_3}\\
16\pi^2\frac{du_2}{dt} &=& u_2\left(2\gamma_{{\cal T}_2^{\rm d}}^{\,\,\,{\cal T}_2^{\rm d}}+\gamma_{{\cal T}_2^{\rm s}}^{\,\,\,{\cal T}_2^{\rm s}}\right)\\
16\pi^2\frac{du_1}{dt} &=& u_1\sum_k\gamma_{{\cal T}_1^k}^{\,\,\,{\cal T}_1^k}\\
16\pi^2\frac{d\beta_i}{dt} &=& \beta_i\left(\gamma_{H_u}^{\,\,\,H_u}+\gamma_{H_d}^{\,\,\,H_d}\right)+\sum_j\beta_j\gamma_{S_j}^{\,\,\,S_i}\\
16\pi^2\frac{d(\lambda_u)_{ij}}{dt} &=& (\lambda_u)_{ij}\gamma_{H_u}^{\,\,\,H_u}+\sum_k(\lambda_u)_{ik}\gamma_{Q_k}^{\,\,\,Q_j}+\sum_k(\lambda_u)_{kj}\gamma_{\bar{U}_k}^{\,\,\,\bar{U}_i}\\
16\pi^2\frac{d(\lambda_d)_{ij}}{dt} &=& (\lambda_d)_{ij}\gamma_{H_d}^{\,\,\,H_d}+\sum_k(\lambda_d)_{ik}\gamma_{Q_k}^{\,\,\,Q_j}+\sum_k(\lambda_d)_{kj}\gamma_{\bar{D}_k}^{\,\,\,\bar{D}_i}\\
16\pi^2\frac{d(\lambda_L)_{ij}}{dt} &=& (\lambda_L)_{ij}\gamma_{H_d}^{\,\,\,H_d}+\sum_k(\lambda_L)_{ik}\gamma_{L_k}^{\,\,\,L_j}+\sum_k(\lambda_L)_{kj}\gamma_{\bar{E}_k}^{\,\,\,\bar{E}_i}
\end{eqnarray}
for the trilinear couplings and
\begin{eqnarray}
16\pi^2\frac{dm_i^k}{dt} &=& m_i^k\left(\gamma_{{\cal T}_i^k}^{\,\,\,{\cal T}_i^k}+\gamma_{\bar{\cal T}_i^k}^{\,\,\,\bar{\cal T}_i^k}\right)\\
16\pi^2\frac{d\mu}{dt} &=& \mu\left(\gamma_{H_u}^{\,\,\,H_u}+\gamma_{H_d}^{\,\,\,H_d}\right)\\
16\pi^2\frac{d\mu_i^2}{dt} &=& \sum_j\mu_j^2\gamma_{S_j}^{\,\,\,S_i}
\end{eqnarray}
for the bilinear and linear couplings.



\end{document}